\documentclass[nofootinbib,superscriptaddress,onecolumn,preprintnumbers,pre]{revtex4}
\newcommand{\mnras}{\textit{MNRAS}}
\newcommand{\jcap}{JCAP}
\newcommand{\physrep}{Phys.\ Rep.}
\usepackage{graphicx}
\usepackage{amsmath,amssymb}
\usepackage{color}

\begin{document}
\title{Complex scalar field in cosmology: axion-like dynamics and deviations from the $\Lambda$CDM model tested with the binned Pantheon Supernova sample}
\author{G. Montani}
\email{giovanni.montani@enea.it}
\affiliation{Nuclear Department, ENEA - C. R. Frascati, Via E. Fermi 45, 00044 Frascati, Italy}
\affiliation{Physics Department, ``Sapienza'' University of Rome,  P.le Aldo Moro 5, 00185 Roma, Italy}

\author{N. Carlevaro}
%\email{nakia.carlevaro@enea.it}
\affiliation{Nuclear Department, ENEA - C. R. Frascati, Via E. Fermi 45, 00044 Frascati, Italy}

\author{M.G. Dainotti}
%\email{maria.dainotti@nao.ac.jp}
\affiliation{Division of Science, National Astronomical Observatory of Japan, 2-21-1 Osawa, Mitaka, Tokyo 181-8588, Japan}
\affiliation{The Graduate University for Advanced Studies (SOKENDAI), Shonankokusaimura, Hayama, Miura District, Kanagawa 240-0115}
\affiliation{Space Science Institute, Boulder, CO, USA}

\author{E. Giovannetti}
\affiliation{Physics Department, ``Sapienza'' University of Rome,  P.le Aldo Moro 5, 00185 Roma, Italy}

\begin{abstract}
We propose a unified description of the dark sector in which dark matter and dark energy correspond, respectively, to the phase and the modulus of a single complex scalar field. The phase behaves as an axion-like field with a QCD-inspired potential, while the modulus evolves near the maximum of a Higgs-like potential, whose value sets the present-day vacuum energy density. This construction naturally couples the two components, so that a dark matter--dark energy interaction could emerge in the late Universe. However, once the dynamics is averaged over the fast axion oscillations, the interaction does not affect the expansion rate. The non-standard scaling of the dark matter density is exactly compensated by its coupling to the modulus, and the residual slow-rolling contribution is negligible. The model is therefore \emph{de facto} indistinguishable from $\Lambda$CDM. To look for deviations, we introduce a phenomenological constant creation rate of axion-like particles. This mechanism yields an effective running Hubble constant that decreases with redshift. Testing it against a binned reconstruction of the Pantheon Type Ia Supernova sample, we find a creation rate of a few percent of the present expansion rate, nonzero at about $2.5\sigma$. The model fits the data better than $\Lambda$CDM and is mildly statistically favored.
\end{abstract}

\maketitle

\section{Introduction}
In the last decade, a clear tension has emerged in the determination of the Hubble constant values from different observational probes \cite{divalentino-review,Santiago2026JHEAp..5100554S,2019NatAs...3..891V,
2021MNRAS.505.3866E, Dainotti2021hubble,2022PhR...984....1S,2022JHEAp..34...49A,
2023Univ....9..393V,2023ARNPS..73..153K,2023Univ....9...94H,Dainotti:2023+t,2025PDU....4901965D,2025MNRAS.543..456C,2025PhRvD.112j3515H,2025PhRvD.112l3509H,2025EPJC...85.1020B,briscese_IJGMMP2026,2025arXiv250524743M,2025arXiv250602235Y,2025arXiv250819770E,2025arXiv250908840L,2025arXiv251114332M,2026A&A...707A.383C,2026Symm...18..207C,2026JCAP...02..043B,2026arXiv260110127L,schiavone-prd-2026,2026arXiv260201543H,2026arXiv260311706K,
Carmesin:2026xua,Krishnan:2020obg,OColgain2025DES5YR}. This inconsistency is particularly significant (more than five $\sigma$) between the measurements obtained by the Planck Collaboration from cosmic microwave background photons \cite{Planck:2018vyg} and those inferred by the SH0ES Collaboration, which adopts a direct distance ladder calibration based on Type Ia Supernovae (SNeIa) \cite{Scolnic2018P,Scolnic:2021amr,Brout:2022vxf}. This relevant discrepancy in the inferred values of the Hubble constant is commonly referred to as the ``Hubble tension'' and has attracted increasing attention in the search for a reliable explanation \cite{2025PDU....4901965D}. The cosmological scenario has become even more puzzling following the data releases of the DESI Collaboration \cite{2025JCAP-DESI,DESI:2025zgx}, which suggest that a Chevallier--Polarski--Linder (CPL) model \cite{Chevallier2001,Linder2003}, also known as the $w_0w_a$ model, is statistically favored over the standard $\Lambda$ Cold Dark Matter ($\Lambda$CDM) model \cite{Weinberg2008} when Baryon Acoustic Oscillation data are analyzed. This result provides a significant indication that dark energy (DE) may not be properly described by a cosmological constant but could instead have an evolving nature (see also \cite{DymnikovaKhlopov1998,DymnikovaKhlopov2001,Mukhopadhyay--1,Doroshkevich-1,DoroshkevichKlypinKhlopov1989}). 

Some authors \cite{Alestas2020PhRvD.101l3516A,Dainotti2021hubble,2022Galax..10...24D,2024arXiv240507039B,2024arXiv240811031P,arx2206.11447,2019MNRAS.483.4803L,2025JHEAp..45..290D,Dainotti2023mnras,2025arXiv250902636L,2021PhRvD.103j3509K,2022arXiv220113384K,Teng2021DarkEnergyEoS} have suggested a possible effective running of the Hubble constant within the same class of sources, specifically when SNIa data have been analyzed through a binned representation of their samples \cite{Dainotti2021hubble}. A possible explanation for such an effective running Hubble constant may arise from purely astrophysical effects, such as a conjectured redshift evolution of SNeIa \cite{2021ApJ...914L..40D}, as implicitly assumed in the phenomenological power-law scaling of the Hubble constant proposed in \cite{Dainotti2021hubble,2022Galax..10...24D}. However, a physical interpretation of this power-law behavior has also been provided in the framework of metric $f(R)$ gravity and in various scenarios involving dark matter (DM)–DE interaction \cite{schiavone_mnras,2024PDU....4601652E}. Although power-law scaling remains, to date, statistically favored in describing binned SNIa data, several alternative theoretical interpretations have been proposed \cite{2024arXiv240415977M,2024PDU....4601652E,deangelis-fr-mnras,fazzari2025,navone-jheap-2026}. These works show that the apparent decreasing trend of the effective Hubble constant can be consistently attributed to different background physical scenarios, all of which are statistically competitive with the standard $\Lambda$CDM model, which instead predicts a constant value across the entire SNIa redshift range.

In this paper, we investigate a specific model of DM-DE interaction \cite{Wang_2016}, as discussed in \cite{2024arXiv240415977M} (see also \cite{deangelis-fr-mnras,fazzari2025}), adopting here a more fundamental quantum field theory approach. In particular, we model the DM contribution through an axion like field, while DE is described by a $\lambda$$\phi^4$ model characterized by a large constant energy density term. A key feature of our framework is that the axion like and DE fields correspond, respectively, to the phase and the modulus of a single complex scalar field (see also the analysis in \cite{mainini07}). As a consequence, an interaction term between the two cosmological components could, in principle, arise. Furthermore, we analyze the evolution of the model near the maximum of the $\lambda$$\phi^4$ potential, which can be associated, without loss of generality, with the present-day Universe. This approach leads to a reduced dynamical description of the late-time cosmological evolution, which we implement within the diagnostic framework of the effective running Hubble constant. The resulting theoretical prediction is then compared with the binned SNIa Pantheon dataset \cite{Dainotti2021hubble}.

As first theoretical result, a careful analysis 
of the dynamics shows how the 
resulting model is, \emph{de facto}, 
indistinguishable from a standard 
$\Lambda$CDM model, so probing the 
conceptual value of the proposed 
unified picture. 

Then, in view of identifying a
possible deviation of the late Universe dynamics from the
$\Lambda$CDM model, we consider
a phenomenological process of
creation of the axion-like particles,
to be interpreted as the net effect of the time variation of the
cosmological gravitational field \cite{parker-1968,calvao-lima-1992} or as
the result of cosmological strings decaying (\cite{2027JHEAp..5500708M} and references therein).
Thus, we introduce in the equation
of the axion scalar field a constant creation rate that induces a
slow decreasing (for increasing redshift values) of the DM contribution in the Universe
Hubble parameter, with respect to
the $\Lambda$CDM dynamics.
In addition to the Hubble constant and
the present-day value of the matter
density parameter, a new free parameter is introduced in the dynamics,
i.e. the constant creation rate.

Adopting the diagnostic tool of
the effective running Hubble constant, we compare the prediction of our
revised cosmological dynamics with
the binned Pantheon sample of the
SNeIa. This model testing
outline that our model is mildly
statistically preferred by data
with respect to the $\Lambda$CDM representation, in which the
effective running Hubble constant
reduces to a real constant.

The paper is organized as follows. In Sec.~\ref{sec2}, we introduce the complex scalar field model: its phase is identified with an axion-like DM component, while its modulus, evolving near the maximum of a Higgs-like potential, accounts for the DE contribution. In Sec.~\ref{sec3}, we derive the corresponding cosmological dynamics in a flat isotropic Universe and show that, once averaged over the fast axion oscillations, the resulting Hubble parameter is \emph{de facto} indistinguishable from the $\Lambda$CDM one. In Sec.~\ref{sec:alp}, we introduce a phenomenological constant creation rate of axion-like particles, which induces small deviations from $\Lambda$CDM, and define the associated effective running Hubble constant. In Sec.~\ref{sec4}, we test the model against the binned Pantheon SNIa data, comparing its statistical performance with the $\Lambda$CDM and power-law scenarios. Concluding remarks follow.

\section{Physical Background}\label{sec2}
In the following analysis, we present a peculiar scenario in the context of DM-DE interaction models \cite{Wang_2016}. In particular, we consider a cosmological dynamics that is characterized by the presence of a complex scalar field, apart from the baryonic matter contribution. The phase of this field is identified with the DM axion contribution, according to the standard shape of its potential term emulated from QCD \cite{KolbTurner1990}. Instead, the modulus of the scalar field is associated with a potential resembling a Higgs profile. The major contribution of this field is in the vacuum energy density of the present Universe, corresponding to the value taken by its potential term near the maximum of the Higgs configuration. We study the dynamics of the late Universe when the complex scalar field configuration has a modulus near such a Higgs maximum and the phase is sufficiently small to attribute a massive character to the corresponding axion-like field. The cosmological dynamics is investigated by averaging over the massive oscillations of the scalar field phase, so that a well-identified DM contribution of the axion-like form is recovered and it turns out as interacting with the Higgs-like field, i.e. a weak DM-DE interaction could emerge in the late Universe.

We consider a spacetime with a metric tensor $g_{\mu\nu}$ ($\mu ,\nu =0,1,2,3$), and a complex scalar field on it, with modulus $\rho$ and phase $\theta$, whose dynamics is described via the Lagrangian density
\begin{equation}
	\mathcal{L} = \frac{1}{2}\partial_{\mu}\rho\partial^{\mu}\rho 
	+\frac{1}{2}\rho^2\partial_{\mu}\theta \partial^{\mu}\theta - V(\rho ,\theta )
	\, , 
	\label{lt1}
\end{equation}
where, the potential term reads as
\begin{equation}
	V = \epsilon_{\Lambda} - \frac{1}{2}\mu_0^2\left(\rho - \sigma_0\right)^2 + \frac{1}{24}\lambda_0 \left( 
	\rho - \sigma_0\right)^4
	+ m_a^2\rho^2 
	\left( 1-\cos \theta  \right)
	\, . 
	\label{lt2}
\end{equation}
Above, $\mu_0$ and $\lambda_0$ are positive constants, while $\sigma_0$ is the energy scale at which the phase field, below interpreted as the axion field, acquires a non-frozen dynamics and it is related to the QCD energy scale $\lambda_{QCD}$ by the relation $\sigma_0\simeq \lambda^2_{QCD}/m_a$, where $m_a$ denotes the axion mass. Finally, we stress that $\epsilon_{\Lambda}$ is the energy density of the potential term for the unstable point $\rho = \sigma_0\, ,\, \theta =0$ and it will be identified with the constant vacuum energy density of the Universe.

The field equations that describe the classical dynamics of $\rho$ and $\theta$ take the form
\begin{equation}
	\nabla_{\mu}\partial^{\mu}\rho 
	- \rho \partial_{\mu}\theta\partial^{\mu}\theta + \partial_{\rho}V = 0
	\, ,
	\label{lt3}
\end{equation}
and
\begin{equation}
	\nabla_{\mu}\left( \rho^2
	\partial^{\mu}\theta\right) + 
	\partial_{\theta}V = 0
	\, , 
	\label{lt4}
\end{equation}
respectively, where above $\nabla_{\mu}$ denotes the covariant derivative with respect to the metric $g_{\mu \nu}$. 

Now, we expand the potential term near $\rho = \sigma_0$ and $\theta = 0$ and, to this end, we introduce the auxiliary field $\delta \rho \equiv \rho - \sigma_0$. We retain terms up to $\delta \rho^2$ and $\theta^2$, so that our potential term takes the form 
\begin{equation}
	V(\delta \rho ,\theta ) \simeq 
	\epsilon_{\Lambda}
	-\frac{1}{2}\mu_0^2\delta \rho^2 
	+ \frac{1}{2} 
	m_a^2\left(\sigma_0+\delta \rho\right)^2\theta ^2
	\, . 
	\label{lt5}
\end{equation}
By means of this interaction term, Eq.(\ref{lt3}) becomes
\begin{equation}
	\nabla_{\mu}\partial^{\mu}\delta\rho - \left(\sigma_0+\delta\rho\right) \partial_{\mu}\theta\partial^{\mu}\theta- \mu_0^2\delta\rho + 
	m_a^2\left(\sigma_0 + \delta \rho\right)\theta^2 = 0
	\, , 
	\label{lt6}
\end{equation}
while Eq.(\ref{lt4}) reads as
\begin{equation}
	\nabla_{\mu}\left[ 
	\left(\sigma_0+\delta\rho\right)^2\partial^{\mu}\theta \right] + 
	m_a^2\left( \sigma_0 +\delta\rho\right)^2\theta = 0
	\, .
	\label{lt7}
\end{equation}
The field equations above are the fundamental dynamical tool at the ground of our cosmological investigation.

\subsection{Motivation for a complex scalar field}
Before addressing the cosmological analysis of the proposed DM-DE interaction model, it is useful to stress some 
fundamental questions at the basis of the complex scalar field formulations and specific features of the field dynamics we will develop in the following. 

As a preliminary consideration, we observe that a complex scalar field endowed with a quadratic potential in the modulus only is, \emph{de facto}, 
entirely equivalent to two independent scalar fields. 
However, this is not the case considered here, and not only because the modulus potential is not quadratic, but also because of the presence of the coupling term $\propto \rho^2(1-\cos\theta )$. 
Thus, the field picture presented here 
amounts to that of two non-linearly interacting scalar fields, 
i.e. two independent but coupled scalar degrees of freedom.

It is worth stressing that the analysis of the classical dynamics of the modulus 
and phase of the complex scalar field is justified 
by the bosonic nature of their 
quantum states, which can reach arbitrarily high occupation numbers, 
in analogy with the behavior of an 
electromagnetic field state.
In particular, the axion-like field, 
to which the phase field reduces in the 
vicinity of the modulus potential maximum, is to be regarded as a Bose condensate \cite{kittle} in which the ground state contains essentially all the 
massive particles and the thermalization is ensured by the axion self-gravity, see the analysis in \cite{sikivie-prl}. 

The choice of a Higgs-like potential for the modulus is rather natural 
in studying symmetry breaking processes and it is crucial in establishing the emergence of a Goldstone boson \cite{coleman}, which can be identified with the phase field. 
Here, we make this choice on the 
basis of this quantum field theory
paradigm and we implement the dynamics very close to the potential maximum. It is precisely this local configuration that allows us to recover the phase field as a real axion-like field, 
ensuring that the phase-modulus coupling term mainly reduces to a mass-like contribution for the phase. 
In this respect, we note that
a large value of the axion mass with respect to the Hubble rate is at the basis of the possibility of introducing an averaging procedure on the phase field, according to the axion proposal in cosmology. Therefore, the form of the 
two dynamical equations for the
$\rho$ and $\theta$ energy densities 
is viewed as the result of the two-field coupling only after 
the average over the fast oscillations of the phase field has been carried out. 

Finally, we would like to emphasize that the choice of a complex field to represent both the dark matter and dark energy contributions is motivated by the idea of a unified picture, in which two fundamental components of the Universe, \emph{de facto} the so-called 
``dark Universe'', emerge from the 
same fundamental physical origin. 
This idea is also corroborated by recent conjectures about a possible interaction between dark matter and dark energy constituents, as 
a natural source of new physics to address tensions in cosmology \cite{2025PDU....4901965D}.
In fact, in the unified picture of a single complex field accounting for both these degrees of freedom, 
their interaction can emerge from the self-interaction of the complex scalar field itself.

\section{Cosmological dynamics}\label{sec3}
We consider a flat isotropic Universe \cite{2020MNRAS.496L..91E} being described by the line element
\begin{equation}
	ds^2 = dt^2 - a^2(t)dl^2
	\, ,
	\label{lt8}
\end{equation}
where $t$ is the synchronous time 
(here we are in natural units $c=\hbar =1$), $dl^2$ the Euclidean line element and $a(t)$ the cosmic scale factor, responsible for the space expansion.

In the proposed model, the Universe is filled by baryonic matter and the complex scalar field only, so that the Friedmann equation takes the following form:
\begin{equation}
	H^2 \equiv \left(\frac{\dot{a}}{a}\right)^2 = \frac{\chi}{3}
	\left( \frac{\epsilon_b^0}{a^3} + \frac{1}{2}\sigma_0^2 \dot{\theta}^2 + \frac{1}{2}\sigma_0^2m_a^2\theta^2 + \epsilon_{\Lambda} \right) + 
	\delta H^2 
	\, , 
	\label{lt9}
\end{equation}
where $\chi$ is the Einstein constant, the dot denotes differentiation with respect to $t$, $\epsilon_b^0$ the today value of the baryonic matter energy density (we fixed to unity the today value of the scale factor) and we set
\begin{equation}
	\delta H^2 \equiv 
	\frac{\chi}{6}\left[
		\dot{\delta \rho}^2 + (2\sigma_0\delta \rho + \delta\rho^2)(\dot{\theta}^2 + m_a^2\theta^2) -\mu_0^2\delta\rho^2  
	\right]
	\, .
	\label{lt10}
\end{equation}
Furthermore, Eq.(\ref{lt6}) takes the homogeneous form
\begin{equation}
	\ddot{\delta \rho} + 3H\dot{\delta \rho} - \mu_0^2\delta \rho 
	+ \left( \sigma_0 + \delta \rho\right)\left(m_a^2\theta^2 - \dot{\theta}^2\right) = 0
	\, , 
	\label{lt11}
\end{equation} 
while Eq.(\ref{lt7}) can be stated as
\begin{equation}
\ddot{\theta} + 3H\dot{\theta} 
+ m_a^2\theta + 2 
	\frac{\dot{\theta}\dot{\delta\rho}}{\sigma_0+\delta\rho } = 0\, .
	\label{lt12}
\end{equation}
The dynamical scheme traced above identifies in the phase $\theta$ the 
axion-like field accounting for the DM contribution, which weakly interacts with the DE component, mainly due to the value of the potential term near the Higgs-like field maximum in $\rho = \sigma_0$.

The field $\theta (t)$ clearly undergoes damped oscillations with a period $\sim m_a^{-1}$ but we are interested in their average over many periods (this situation is due to the much larger value of the axion mass $m_a\sim[10^{-6},10^{-3}]$ eV compared to the Hubble mass $m_H\equiv H_0\sim 10^{-33}$ eV). In the first approximation, we can neglect the term in $\dot{\delta\rho}$ of Eq.(\ref{lt12}) (i.e. $\dot{\delta\rho}/\sigma_0\ll H$), so that we get the following leading order solution
\begin{equation}
	\theta (t)\simeq 
	\frac{\theta_0}{a^{3/2}}\sin (m_at + \theta^*) \rightarrow 
	\epsilon_a \equiv \frac{1}{2}\sigma_0^2\dot{\theta}^2 + \frac{1}{2}\sigma_0^2m_a^2\theta^2 
	\simeq \sigma_0^2\langle \dot{\theta}^2\rangle 
	\simeq \sigma_0^2m_a^2\langle \theta^2\rangle
	\, 
	\label{lt16}\,,
\end{equation}
in which $\theta_0$ and $\theta^*$ 
are integration constants and $\langle ...\rangle$ denotes average over many oscillation periods, which are many orders of magnitude faster than the time scale $H^{-1}$. By means of Eq.(\ref{lt16}), Eqs.(\ref{lt11}) and (\ref{lt12}) rewrite as
\begin{eqnarray}
	\ddot{\delta\rho} + 
    	3H\dot{\delta\rho} 
	-\mu_0^2\delta\rho = 0
	\, , \label{feq1}\\
	\dot{\epsilon}_a = - \Big( 
	3H + 2\frac{\dot{\delta\rho}}{\sigma_0+\delta\rho }\Big) \epsilon_a 
	\, . \label{feq2}
\end{eqnarray}
We stress that in the last equation we retain the term $\dot{\delta\rho}$ because this dynamics is now averaged over the small oscillations. Using the redshift time variable $z$, it is immediate to recognize that 
Eq.(\ref{feq2}) admits the solution
\begin{equation}
	\epsilon_a = \epsilon_a^0(1+z)^3\Big( 1 + \frac{\delta\rho}{\sigma_0}\Big)^{-2} 
	\, , 
	\label{lt19}
\end{equation}
where $\epsilon_a^0$ is an integration constant.

Hence, by means of the solution above, the Friedmann equation in Eq.(\ref{lt9}) and Eq.(\ref{lt10}),   
once averaged over the $\theta$-oscillations, can be conveniently rewritten as follows:
\begin{equation}
	H^2=H_0^2\Big( 
	\Omega_m^0(1+z)^3 
	+ 1-\Omega_m^0 - 
	\frac{\chi\mu_0^2}{6H_0^2}(\delta\rho^2-\delta\rho^2(z=0))\Big) 
	\,.
	\label{lt21}
\end{equation} 
Here, we introduce the critical density parameters $\Omega_i^0\equiv \chi \epsilon_i^0/3H_0^2$ (for $i=a,b,\Lambda$ and $\epsilon_\Lambda^0\equiv\epsilon_\Lambda$), and we use the normalization condition $\Omega_{\Lambda}\equiv 1-\Omega_m^0+\chi\mu_0^2\delta\rho(z=0)^2/6H_0^2$, with $\Omega_m^0\equiv \Omega_a^0 + \Omega_b^0$, since $H_0\equiv H(z=0)$. 

In order to understand the range of the basic parameters of the model, it is worth constructing normalized equations. To this end, we introduce the dimensionless quantities: 
\begin{equation}
	\xi \equiv \frac{\delta\rho}{\sigma_0}\, ,\qquad
   \qquad
	\sqrt{\chi}\sigma_0 = 
	\frac{\lambda_{QCD}^2}{m_am_{Pl}}
	\, , 
	\label{lt22}
\end{equation}
where $m_{Pl}$ denotes the Planck mass. Adopting the reasonable values $\lambda_{QCD}\sim 200$ MeV, $m_a \sim[10^{-6},10^{-3}]$ eV and recalling that $m_{Pl}\sim 10^{18}$ GeV, we arrive to the following identification: $\sqrt{\chi}\sigma_0 \sim [10^{-8},10^{-5}]$, where the indicated ranges correspond to the interval for the axion mass $m_a$. Since we set the present-day value of the scale factor equal to unity, from Eq.(\ref{lt16}) we see that $\theta_0$ is essentially the today value of $\theta(t)$. It is important to stress that the experimental limits on the value of $\theta_0$ state that its value should be below $10^{-10}$ \cite{Kim:2008hd,Peccei_2008}.

In order to implement a slow-rolling picture for the scalar field 
$\delta \rho$, we have to require that the maximum of the potential profile be sufficiently flat, i.e. we have 
to deal with a sufficiently small 
value of the parameter $\mu_0$. 
In particular, the proper requirement stands as $\mu_0\ll H_0$, recalling that we are considering the late cosmological dynamics. 
Therefore, under this hypothesis, 
Eq.(\ref{feq1}), once expressed 
in the redshift time variable, takes 
the following form: 
\begin{equation}
	(1+z)\frac{d\xi}{dz} 
	= - \bar{\mu}_0^2 
	\frac{\xi}{E^2}
	\, , 
	\label{xi1}
\end{equation}
where $\bar{\mu}_0^2\equiv 
\mu_0^2/3H_0^2\ll 1$ 
and $E\equiv H/H_0$ is the Universe 
expansion rate. Easily solving 
the equation above, we can restate the Friedmann equation (\ref{lt21}) 
in its final form as
\begin{equation}
	E^2 = \Omega_m^0 
	(1+z)^3 + 1 - \Omega_m^0 +  \Omega_{\mu}^0\xi_0^2\Big[ 
	1-\exp\Big(-2\bar{\mu}_0^2
	\int_0^z\frac{dy}{(1+y)E^2(y)}\Big)\Big]
	\, , 
	\label{ixi2}
\end{equation}
where we have introduced $\Omega_{\mu}^0\equiv 
\chi\mu_0^2\sigma_0^2/6H_0^2$, 
while $\xi_0\equiv \xi(z=0)\ll1$.
Since, in the best case, we have 
$\sqrt{\chi}\sigma_0 \simeq 
\sigma_0/m_{Pl}\simeq 10^{-5}$ and 
taking $\bar{\mu}_0\simeq 10^{-2}$, 
we easily conclude that 
$\Omega_{\mu}^0<10^{-14}$. 
By other words, the present model is today \emph{de facto} indistinguishable from a $\Lambda$CDM dynamics.

Actually, the conclusion that the 
scenario proposed here reduces 
essentially to the Hubble parameter of 
a standard $\Lambda$CDM model, although 
the DM component apparently scales with 
the slightly modified law in Eq.(\ref{lt19}), can be viewed as a general statement under the following 
considerations. In this unified picture, the 
axion-like DM contribution is naturally coupled to the squared 
modulus, but, after averaging over the rapid small oscillations of the 
phase around the vacuum, the associated energy density has an inverse squared dependence on the modulus, as 
in Eq.(\ref{lt19}) (valid also in a non-perturbative approach) and the net 
effect of this compensation is that, in the Hubble parameter, $\Omega_a$ 
appears in its standard form. Moreover, in this picture, we introduce a self-interacting scalar field contribution, corresponding to the modulus, i.e. a quintessence component. But, near the maximum of 
the potential for the modulus field 
(in principle also near a minimum), 
the dynamics of this quintessence 
is naturally associated to a slow-rolling phase (but even different situations would not essentially change 
the following statement), in 
which a cosmological constant term 
emerges, apart from a residual negligible dynamics.

\section{Axion-like particle generation}\label{sec:alp}
In view of searching possible modifications of the $\Lambda$CDM model as suggested by recent observations, we modify the picture proposed above by phenomenologically altering the dynamics of the phase field. In this respect, we focus our 
attention on a possible generation mechanism for the axion-like particles, 
according to the time variation of 
the gravitational field in the 
expanding Universe \cite{parker-1968,calvao-lima-1992,2001CQGra..18..193M,schiavone-prd-2026,navone-jheap-2026,odintsov-matter-creation},  
or due to 
the decaying process of cosmological strings (see \cite{2027JHEAp..5500708M} and references therein). We consider a phenomenological constant creation rate 
$\Gamma$ of axion-like particles, 
which allows to replace Eq.(\ref{lt12}) with
\begin{equation}
	\ddot{\theta} + (3H 
	- \Gamma )\dot{\theta} 
+ m_a^2\theta + 2 
	\frac{\dot{\theta}\dot{\delta\rho}}{\sigma_0+\delta\rho } = 0
	\, .
	\label{lt12n}
\end{equation}
Now, in the leading orders, the solution of the equation above stands as follows:
\begin{equation}
	\theta (t) = 
	\frac{\theta_0}{a^{3/2}}
	e^{\Gamma t/2}\sin 
	\left( m_at + \theta ^* 
	\right)
	\, . 
	\label{ixp22}
\end{equation}

Once averaged over the fast axion 
field oscillations (we require 
that $\Gamma \lesssim H \ll m_a$), Eq.(\ref{lt12n}) takes the form
\begin{equation}
	\dot{\epsilon}_a =   
	- \Big( 3H - \Gamma 
+ 2\frac{\dot{\delta\rho}}{\sigma_0 + \delta\rho}\Big) \epsilon_a
\, , 
	\label{ixp23}
\end{equation}
which can be set, using the redshift time variable, in the following 
dimensionless dynamical equation:
\begin{equation}
	\frac{d\Omega_a}{dz} = 
	\frac{\Omega_a}{1+z}
	\left( 3 - \frac{\gamma}{E}\right) - \frac{2}{1+\xi}\frac{d\xi}{dz}
\Omega_a	\, , 
	\label{ixp24}
\end{equation}
where $\gamma \equiv \Gamma /H_0$. As already noted, since $\xi$ decays from $\xi_0 \ll 1$, we can safely neglect it within the dynamical scheme also due to the negligible contribution of $\Omega_{\mu}^0$. Introducing the quantity $\Omega_a^{\textrm{eff}}\equiv\Omega_a(1+z)^{-3}$, Eq.(\ref{ixp24}) finally reads
\begin{equation}
	\frac{d\Omega_a^{\textrm{eff}}}{dz} = 
	-\frac{\gamma}{E}\,\frac{\Omega_a^{\textrm{eff}}}{1+z}\; , 
	\label{ixp24eff}
\end{equation}
with now $\Omega_a^{\textrm{eff}}(0)=\Omega_a(0)\equiv \Omega_a^0$ (since, neglecting $\xi_0$, we have $\epsilon_a^0=\epsilon_a(0)$).
Under the same assumptions, the Friedmann equation 
(\ref{lt21}) takes the new 
dimensionless form
\begin{equation}
	E^2 = \left( \Omega_b^0 
	+ \Omega_a^{\textrm{eff}}(z)
	\right) (1+z)^3 
	+ 1 -\Omega_b^0 - \Omega_a^0
	\, , 
	\label{ixp25}
\end{equation}
recalling that $\Omega_b^0+\Omega_a^0=\Omega_m^0$. Thus, the proposed model contains 
three free parameters, i.e. 
$H_0$, $\Omega_a^0$ and $\gamma$ ($\Omega_b^0$ being fixed by the primordial chemosynthesis).

It is worth noting that our phase field is, in the end, a cold DM candidate, according to the same axion paradigm \cite{KolbTurner1990} and, due to the cancellation described above, the corresponding energy
density scales as $(1+z)^3$,
exactly like the baryonic matter contribution. Thus, from the point of view
of data analysis, these two terms are, in practice, indistinguishable,
but it is worth remarking that the value of the baryonic critical density
parameter is well constrained by
the primordial Big Bang nucleosynthesis
of light elements \cite{Gamow1946,Planck:2018vyg,Steigman2007,Cyburt2016,Pitrou2018,Fields2020}.
This feature is common to other proposals for the dark matter component,
since the data are essentially able to discriminate only the coefficient
of the global term
$\propto (1+z)^3$.

Finally, in order to compare the present model with the binned data of the SNeIa, we introduce the effective running Hubble constant \cite{2022Galax..10...24D,Dainotti2021hubble} (see also \cite{fazzari2025}), defined as
\begin{equation}
\mathcal{H}_0(z) = H_0 
\frac{E(z)}{\sqrt{\Omega_m^0(1+z)^3 + 1-\Omega_m^0}} 
\, , 
\label{lteff}
\end{equation}
where $E(z)$ is taken from Eq.(\ref{ixp25}).

\section{Data analysis}\label{sec4}
In this section, we summarize the data analysis strategy introduced in \cite{Dainotti2021hubble,2022Galax..10...24D} to construct a binned representation of the SNIa Pantheon sample. Throughout the analysis, the Hubble constant $H_0$ is expressed in units of km s$^{-1}$ Mpc$^{-1}$. The dataset adopted here is the Pantheon compilation \cite{Scolnic2018P}, which includes 1048 Type Ia Supernovae collected from multiple observational surveys. The SNe Ia spans up to redshifts 2.26 \citep{Rodney2016} and they are an extension of the JLA sample \citep{Betoule14}. We have chosen the Pantheon sample instead of the Master sample \citep{Dainotti2025JHEAp..4800405D} as we wish to compare this model with other ones in which the Pantheon sample has been used. As a future plan, we will update the Master sample and use it as comparison.

To perform the analysis, the full dataset is partitioned into 40 redshift bins, each containing the same number of supernovae. For every bin, we compute the average redshift, so that the representative point corresponds to the mean value of the SNeIa within that bin. We here note that the interval of the bin is treated as a bin and not as an uncertainty, thus the fitting includes the values of $H_0$ in each bin with their relative uncertainties. Due to the reduced number of objects at higher redshift, the central values of the bins tend to be slightly shifted toward lower $z$, since this binning procedure is equi-populated one. For a different binning procedure method such as in $\log z$ see \cite{dainotti-master-sample} or $\log 10 (1/(1+z))$ see \citep{Dainotti2024PDU....4401428D,Dainotti2024Galax..12....4D}. We here use for the binning analysis Gaussian likelihoods, but for a different treatment of non-Gaussianity likelihoods see \citep{dainotti-jheap-2024}. The observed distance modulus is defined as $\mu_{\text{obs}} = m_B - M$, where $m_B$ denotes the apparent B-band magnitude (including both statistical and systematic corrections) and $M$ represents the absolute magnitude of a reference supernova, corrected for stretch and color effects. Following previous works, the averaging of the distance modulus is performed using the prescriptions of \cite{2010A&A...523A...7G} (G2010) and \cite{2011A&A...529L...4C} (C2011). The theoretical counterpart is given by $ \mu_{\text{th}} = 5 \log_{10} d_L(z, H_0, ...) + 25$, where the luminosity distance $d_L$ (expressed in Mpc) is computed within a fiducial $\Lambda$CDM cosmological model. Corrections associated with the peculiar velocities of the host galaxies are also included. To assess the goodness of fit, we introduce the chi-squared estimator for the SNIa dataset: $\chi^2_{\text{SN}} = \Delta\mu^{T} \cdot \mathcal{C}^{-1} \cdot \Delta\mu$, with $\Delta\mu = \mu_{\text{obs}} - \mu_{\text{th}}$ and $\mathcal{C}$ being the $1048 \times 1048$ covariance matrix provided in \cite{Scolnic2018P}.
We here stress that the $\mathcal{C}$ matrix includes both statistical and systematic uncertainties. 
Within each redshift bin, the value of $H_0$ is inferred by allowing it to vary while keeping $\Omega_{m}^{0}$ fixed at the fiducial value found in \cite{Scolnic2018P}. Parameter estimation is carried out using an MCMC procedure, which provides the best-fit value of $H_0$ in each bin. Concerning the choice of $\Omega_{m}^{0}$, its possible variation across bins is constrained to remain within 2$\sigma$ of the prior values as it is demonstrated in \cite{Dainotti2021hubble}. The selection of 40 bins represents a compromise between statistical robustness and resolution: each bin contains approximately 26 supernovae, ensuring sufficient statistics while avoiding excessive fragmentation of the dataset. Importantly, the reconstructed trend of $H_0$ is insensitive to the initial choice of its value, since the Pantheon distance moduli are calibrated by assuming a fixed absolute magnitude $M$. In particular, we adopt $M = -19.245$, corresponding to $H_0 = 73.5$, as in \cite{Dainotti2021hubble}. The procedure consists in fixing $H_0$ in the first bin, determining $M$ as the only free parameter, and then keeping this value of $M$ fixed for all subsequent bins. This approach does not alter the observed decreasing behavior of $H_0(z)$. The independence of the results from the number of bins is also shown in \cite{2022Galax..10...24D,dainotti-master-sample}. This analysis has been done with Gaussian likelihood following the standard procedure done by \cite{Scolnic2018P}. However, we point out that the difference between the distance moduli theoretical and observed corrected by the covariance matrix, see Figure 1 in \cite{dainotti-jheap-2024}, in the $\Lambda$CDM model is not Gaussian and follows the T-student distribution according to the standard kurtosis, skewness and skewness+kurtosis test. Moreover, neither the choice of binning nor fixing $\Omega_m^0$ affects the results, as shown in Figure 2 of \cite{Dainotti2021hubble}. This analysis also has been done for the Master sample, see \cite{dainotti-master-sample,2026arXiv260300497D}. The reason why we did not use modified likelihood is that we would like to compare the model of standard $\Lambda$CDM with other major work in the literature. In addition, a full procedure would require the fitting of each distribution again in each bin, which goes beyond the analysis needed for this paper.

It is worth emphasizing that, while the SH0ES collaboration extracts a single value of $H_0$ from SNIa data, the analyses presented in \cite{Dainotti2021hubble,2022Galax..10...24D} show that a binning in redshift unveils a gradual decline of $H_0$ with increasing $z$. If the Hubble tension originates from new physics, as hypothesized here, such an effect should manifest continuously across redshift, making an evolving $H_0(z)$ a natural expectation.

We underline that $\mathcal{H}_0(z)$, defined in Eq.(\ref{lteff}), is introduced as a diagnostic tool to test whether a dataset is consistent with the standard $\Lambda$CDM prediction or instead requires deviations described within the given theoretical framework. In general, this function depends on three free parameters, namely $H_0$, $\Omega_a^0$ (or $\Omega_{m}^{0}$) and $\gamma$, all of which could, in principle, be included in the fitting procedure. However, within the binned approach adopted here, we assign fiducial values to $H_0$ and $\Omega_{m}^{0}$, namely $H_0=73.5$ and $\Omega_{m}^{0}=0.298$. We also remark that we assume to take $\omega_b = 0.0224\pm0.0001$ \cite{Planck:2018vyg} (center value), thus we obtain $\Omega_b^0=\omega_b/(H_0/100)^2=0.0415$  and $\Omega_a^0=0.256$.
\begin{figure}[ht!]
\includegraphics[width=0.55\textwidth]{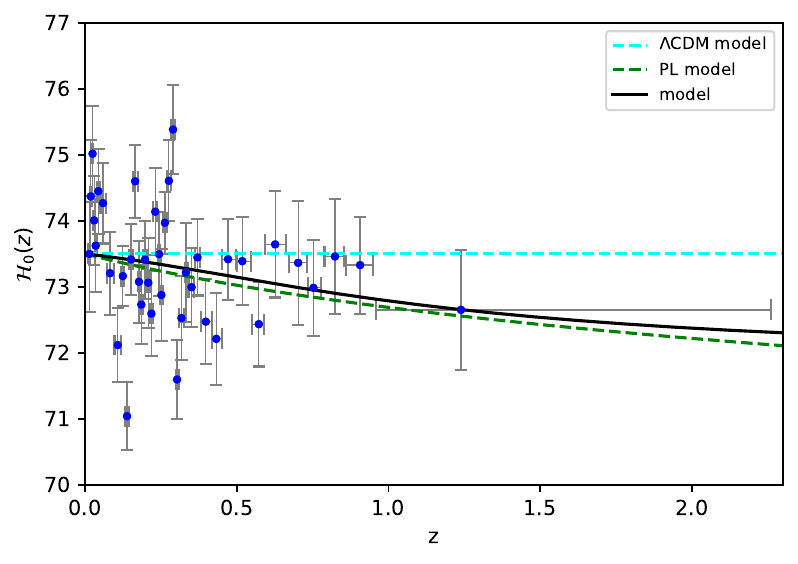}
\caption{Plot of $\mathcal{H}_0(z)$ from Eq.(\ref{lteff}) (black) for the best fit $\gamma$ in Eq.(\ref{bestfit}) and the fiducial values $H_0=73.5$, $\Omega_{m}^{0}=0.298$. The green dashed line represents the profile $H_0(1+z)^{-0.016}$ (PL model). Blue bullets are $H_0$ data from \cite{Dainotti2021hubble} with the corresponding error bars in 1$\sigma$ and bin widths in the $x$-axis. We also depict the constant line $\mathcal{H}_0(z)=73.5$ for the base $\Lambda$CDM model (cyan dashed).}
\label{fig-H0}
\end{figure}

Using the 40-bin reconstruction of $H_0(z)$ described above, we perform a nonlinear fit, obtaining the best-fit value
\begin{align}\label{bestfit}
\gamma=0.053 \pm 0.021\;.
\end{align}
For comparison, we also consider the power-law parametrization $\mathcal{H}_0(z)=H_0 (1+z)^{-\alpha}$, with best-fit value $\alpha=0.016\pm0.009$ (hereafter PL model), as reported in \cite{Dainotti2021hubble}. In addition, when referring to the $\Lambda$CDM scenario we assume a constant effective Hubble parameter $\mathcal{H}_0(z)=H_0=73.5$. In Fig.\ref{fig-H0}, we display the behavior of $\mathcal{H}_0(z)$ for the three models, together with the binned observational data. 

Regarding the statistical performance, we obtain $\chi_{\text{red}}^2=2.0129\, \text{(PL)}, 2.0194\, \text{(model)}, 2.1214\, (\Lambda\text{CDM})$. The large number of data points at low redshift, combined with the high number of degrees of freedom, leads to (inverse) p-values close to unity for all models. The ordering of the $\chi_{\text{red}}^2$ values indicates a slightly better performance of the PL model in reproducing the binned data. The PL behavior was originally motivated by the hypothesis of an intrinsic redshift evolution of SNeIa \cite{2021ApJ...914L..40D} and it is mutuated by a discussion of the evolution in intrinsic properties as the ones we see in Gamma-Ray Bursts \cite{Dainotti_2020,2021ApJ...914L..40D,Bargiacchi2023MNRAS.521.3909B,Dainotti2022PASJ...74.1095D,Dainotti2023ApJ...951...63D} and Quasars \cite{Dainotti2023ApJ...950...45D,Lenart2023}. Furthermore, a theoretical interpretation of such a scaling has been proposed in \cite{schiavone_mnras}, where it emerges within a viable $f(R)$ gravity framework formulated in the Jordan frame.
To compare models with different degrees of freedom, we report the Akaike and Bayesian information criteria: AIC$=120.4265\, \text{(PL)},\, 120.6776\, \text{(model)},\, 124.7775\, (\Lambda\text{CDM})$ and
BIC$=122.1154 \,\text{(PL)},\,122.3665 \,\text{(model)},\, 124.7775\, (\Lambda\text{CDM})$. Taking $\Lambda$CDM as the reference, our model yields
$\Delta\mathrm{AIC}=4.10$
and $\Delta\mathrm{BIC}=2.41$. According to the AIC, our model is therefore
moderately favoured over $\Lambda$CDM, whose relative likelihood is
$e^{-\Delta\mathrm{AIC}/2}\simeq0.13$. The BIC, which penalises the additional
parameter more strongly, provides positive but weak evidence in the same
direction. Nonetheless, the mild preference for the alternative behaviors point towards interesting scenarios beyond the standard cosmological framework, motivating further investigation with future data.

\section{Concluding remarks}
We have proposed a unified description of the dark sector in which DM and DE are associated, respectively, with the phase and the modulus of a single complex scalar field. The phase plays the role of an axion-like field, with a QCD-inspired potential, while the modulus evolves near the maximum of a Higgs-like potential, whose value provides the present-day vacuum energy density. This construction could contain a coupling between the two components, so that a DM-DE interaction emerges in the late Universe.

As a first result, we have shown that, once the dynamics is averaged over the fast oscillations of the axion-like field, this interaction does not affect the Hubble parameter. The DM energy density acquires a non-standard dependence on the modulus, but this is exactly compensated by its coupling to the squared modulus, so that axion-like density parameter enters the Friedmann equation in its standard form. Moreover, in the slow-rolling regime required near the maximum of the potential, the residual contribution of the modulus is suppressed. The model is therefore \emph{de facto} indistinguishable from $\Lambda$CDM. This conclusion holds beyond the specific perturbative scheme adopted: within this unified picture, appreciable deviations from $\Lambda$CDM require an additional, phenomenological, energy exchange between the two components.

We have then introduced a constant creation rate $\Gamma$ of axion-like particles, interpreted as the net effect of particle production by the time-varying cosmological gravitational field or of the decay of cosmological strings. This mechanism induces a slow decrease of the effective DM contribution with increasing redshift, relative to $\Lambda$CDM scheme. The resulting effective running Hubble constant $\mathcal{H}_0(z)$ decreases with $z$. Fitting it to the 40-bin reconstruction of $H_0(z)$ from the Pantheon sample, we obtain $\gamma=\Gamma/H_0=0.053\pm0.021$. This value is nonzero at about $2.5\sigma$ and corresponds to a creation rate of a few percent of the present expansion rate, consistent with the assumption $\Gamma\sim H\ll m_a$.

From a statistical point of view, the model improves the description of the binned data with respect to $\Lambda$CDM, with a reduced $\chi^2$ close to that of the phenomenological power-law scaling. It is also mildly favored by the AIC. The BIC, however, which penalizes the additional parameter more strongly, indicates that the three scenarios remain statistically indistinguishable with the current dataset. Thus, the present analysis shows that axion-like particle creation in a unified dark sector provides a physically motivated alternative to the $\Lambda$CDM model.

\section*{Acknowledgment}

We would like to thank Ramzi Ziriat for the interesting discussions on the validity of the approximation performed on the proposed dynamical system. M.G.D. acknowledges the support of the JSPS Grant-in-Aid for Scientific Research (KAKENHI) (A), Grant Number JP25H00675.
    
%\bibliographystyle{biblio_style}
%\bibliography{biblio_astro}

\end{document}